\documentclass[aps,pre,preprint,superscriptaddress,onecolumn,showkeys]{revtex4-2}

\usepackage{color}

\usepackage{epsfig,amssymb,amsmath,amsthm,amsfonts,amsbsy,mathrsfs}

\usepackage{graphicx}

\usepackage{subfigure}

\usepackage{float}

\usepackage{verbatim}

\usepackage{physics}

\usepackage{epstopdf}

\newcommand{\bx}{{\bf x}}
\newcommand{\drmt}{\hspace{2pt}\Delta t}

\begin{document}

\title{Breakdown of Boltzmann-type Models for the Alignment of Self-propelled Rods}

\author{Patrick Murphy\footnote[1]{These authors contributed equally to this project}
\footnote[2]{corresponding author}
}
\affiliation{Department of Bioengineering, Rice University, Houston, TX 77005}
\affiliation{Center for Theoretical Biological Physics, Rice University, Houston, TX 77005}

\author{Misha Perepelitsa$^{\textrm{a}}$
}
\affiliation{Department of Mathematics, University of Houston, TX 77204}
\email[Corresponding author: ]{mperepel$@$central.uh.edu}

\author{Ilya Timofeyev}
\affiliation{Department of Mathematics, University of Houston, TX 77204},

\author{Matan Lieber-Kotz}
\affiliation{Department of Bioengineering, Rice University, Houston, TX 77005}

\author{Brandon Islas}
\affiliation{Department of Computational and Applied Mathematics, Rice University, Houston, TX 77005}

\author{Oleg A. Igoshin$^\textrm{b,}$}
\affiliation{Department of Bioengineering, Rice University, Houston, TX 77005}
\affiliation{Center for Theoretical Biological Physics, Rice University, Houston, TX 77005} 
\affiliation{Department of Chemistry, Rice University, Houston, TX 77005}
\affiliation{Department of Biosciences, Rice University, Houston, TX 77005}
\email[Corresponding author: ]{igoshin$@$rice.edu}
\date{\today}

\begin{abstract}

Studies in the collective motility of organisms use a range of analytical approaches to formulate continuous kinetic models of collective dynamics from rules or equations describing agent interactions. However, the derivation of these kinetic models often relies on Boltzmann's hypothesis of "molecular chaos", that correlations between individuals are short-lived. While this assumption is often the simplest way to derive tractable models, it is often not valid in practice due to the high levels of cooperation and self-organization present in biological systems. In this work, we illustrated this point by considering a general Boltzmann-type kinetic model for the alignment of self-propelled rods where rod reorientation occurs upon binary collisions. We examine the accuracy of the kinetic model by comparing numerical solutions of the continuous equations to an agent-based model that implements the underlying rules governing microscopic alignment. Even for the simplest case considered, our comparison demonstrates that the kinetic model fails to replicate the discrete dynamics due to the formation of rod clusters that violate statistical independence. Additionally, we show that introducing noise to limit cluster formation helps improve the agreement between the analytical model and agent simulations but does not restore agreement completely. These results highlight the need to both develop and disseminate improved moment-closure methods for modeling biological and active matter systems.

\end{abstract}


\maketitle


\section{Introduction}

\begin{figure}[htbp]
\centering
\includegraphics[clip, trim=0cm 1cm 0cm 2cm, width=1.00\textwidth]{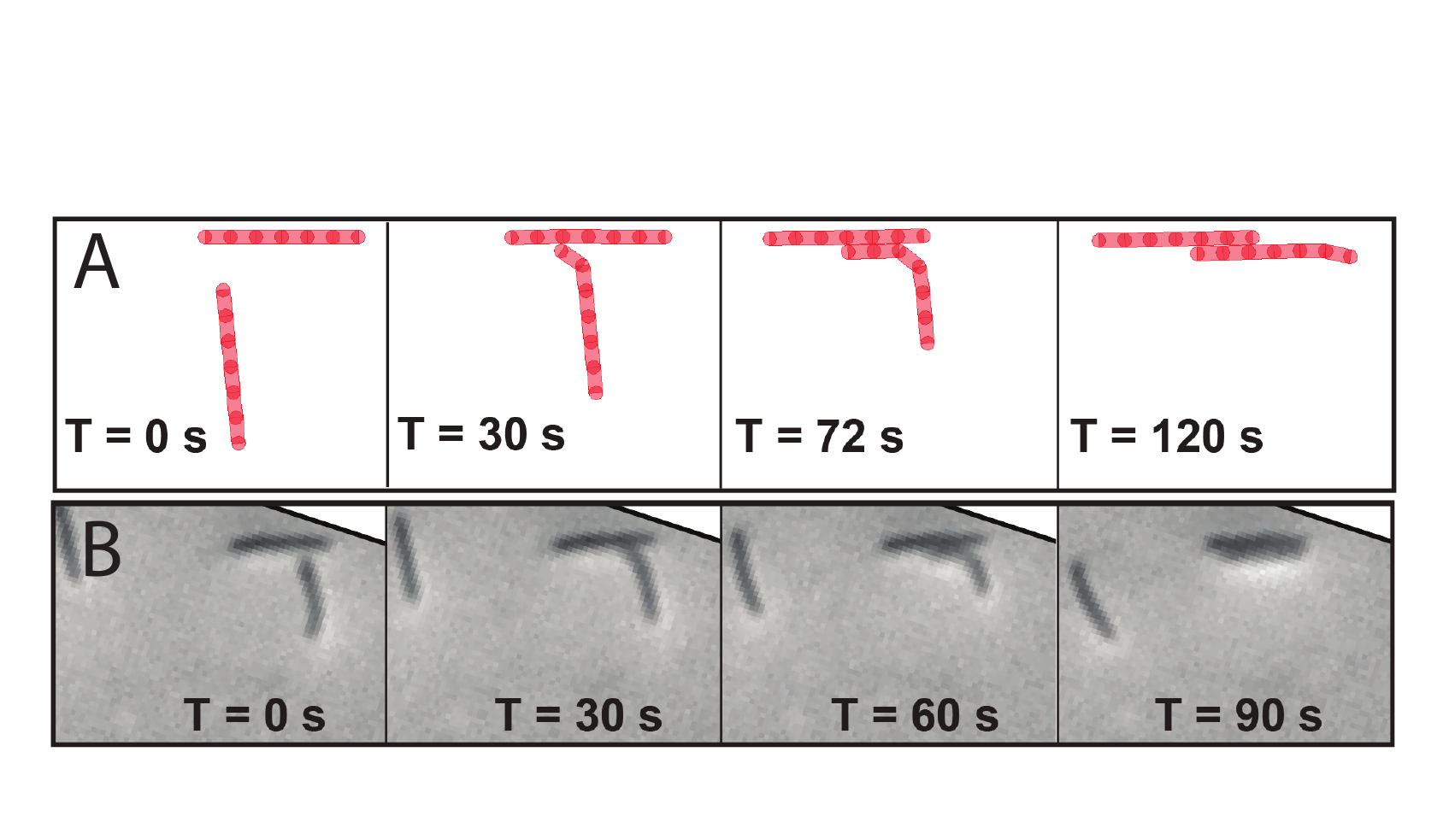}

\caption{An example of asymmetric alignment during the collision of two \textit{M. xanthus} cells due to cell-substrate forces. Adapted from \cite{balagam14b}.
\label{fig:myxo}
}
\end{figure}

Self-propelled rods are a fascinating class of active matter seen across both living and non-living systems \cite{ndlec97,riedel05,narayan07,schaller10,sumino12,balagam14b,balagam15b}. Due to their shape, such systems are intrinsically capable of collective behavior through realignments of rods due to physical collisions \cite{baskaran08,baskaran08a} or longer-ranged hydrodynamic interactions when in a fluid \cite{riedel05,baskaran09}.
These interactions lead to the emergence of macroscopic collective motion such as flocking, clustering, phase changes, and vortexes. Two notable biological examples of collective motion in self-propelled rods are the dynamics of rod-shaped gliding bacteria such as the soil bacterium \textit{Myxococcus xanthus} \cite{kaiser03,wu07,thutupalli15c}, and the behavior of groups of cellular cytoskeletal rods (such as F-actin \cite{schaller10} or microtubules \cite{sumino12}) driven by molecular motors deposited on the surface. Collisions between \textit{M. xanthus} cells often result in the head of the colliding cell reorienting along the length of the cell that was struck (see Fig. \ref{fig:myxo}) \cite{balagam14b}. Reorientation itself occurs on a characteristic time scale related to the ratio of the cell length to cell velocity. Microtubule collisions on a 2D surface also exhibit asymmetric realignments upon collision, with the additional possibility of the colliding microtubule stalling until the struck microtubule has passed \cite{sumino12}.

Due to the interest surrounding systems of self-propelled rods, researchers have developed various methods of modeling them. A popular approach to studying the emergent behavior of these systems is by deriving a probabilistic description \cite{Bertin,degond18,degond08,Manhart1,baskaran08a,peruani06a,ihle_kinetic_2011,bolley12,hanke_understanding_2013,peshkov_boltzmann-ginzburg-landau_2014,perepelitsa_mean-field_2022}. 
A number of heuristic mean-field models based on the notion of the mean nematic director have been considered in \cite{Peruani2008, Ginelli2010, degond08, degond10, degond17, degond18}. Since rod reorientation takes place only when rods collide, a more accurate approach consists in tracing rod collisions as they naturally occur and updating the kinetic distribution function accordingly.

A common starting point for such derivations is an application of Boltzmann framework that originated in the kinetic theory of gases. The framework is based on 4 key elements: i) the geometry (shape) of interacting particles; ii) the type of interactions, that determines how particle states change in collisions; iii) the proper asymptotic regime in the model parameters; and iv) the assumption of ``molecular chaos", that is the absence of two-particle correlations. Importantly, the validity of hypothesis iv) depends on the geometry of interacting particles and the type of interactions.

This modeling approach for self-propelled particles aligning through collisions has been taken up by several authors \cite{Bertin,Manhart1,hanke_understanding_2013,perepelitsa_mean-field_2022,thuroff_critical_2013,ihle_towards_2014,suzuki_polar_2015}.
In Bertin et al (2006) \cite{Bertin}, authors consider spherically shaped cells that experience {\it polar alignment} with some amount of noise in the post-collisional orientations. Their model is based on the zero-order approximation of the interaction operator, which amounts to treating terms of order $\frac{Nl^2}{L^2}$ as negligible,  where $N$ represents the number of particles, $l$ denotes the particle size (radius), and $L$ signifies a macroscopic length scale. This approximation signifies an asymptotic regime where variations in the kinetic function across distances comparable to the particle size can be disregarded, reminiscent of the classical Boltzmann equation.
Under these conditions and hypothesis iv), the authors derived a kinetic equation for the distribution of cells and a reduced hydrodynamic-type model for the first few moments of the cell density distribution, assuming that the macroscopic motion is slow.

Cells of more realistic shape (rod-shaped cells) were considered in Hittmeir et al (2021) \cite{Manhart1}. The authors built a kinetic model in which a pair of colliding co-oriented cells (rods with angle difference $|\theta_1-\theta_2| < \pi/2$) change their orientation to their average alignment;  otherwise ($|\theta_1-\theta_2| > \pi/2$) both colliding cells reverse their direction of motion. This complex interaction consists of a partial {\it polar alignment} and a partial reversal. Again, the kinetic model is obtained using the zero-order approximation of the interaction operator and the molecular chaos assumption. For special sets of cell orientations, the authors identify the corresponding set of equilibrium distributions and derive a hydrodynamic-type model in the limit $\frac{Nl}{L}\to\infty.$

Over the last decade, the validity of Boltzmann formalism has come into question. Evidence has shown that in some biological systems, this modeling approach is not enough to reproduce the observed system dynamics due to either weak binary interactions when stronger, multiparticle interactions are needed \cite{suzuki_polar_2015}, or due to rapid cluster formation that violates the molecular chaos hypothesis \cite{hanke_understanding_2013,thuroff_critical_2013,ihle_towards_2014}. These latter studies suggest that biologically-relevant phenomena reliant on cluster formation, such as the transitions from unordered to collective motion \cite{peruani06a,peruani_collective_2012,weber_nucleation-induced_2012}, may not be accurately modeled with Boltzmann-type equations. However, results from Thüroff et al (2013) \cite{thuroff_critical_2013} suggest that a Boltzmann approach can be appropriate for systems that align weakly or gradually over time.


In this paper, we present a refined kinetic model that offers a more precise representation of cell interactions, taking into account both the nature of these interactions and the specific asymptotic regime governing the interaction operator. Furthermore, we examine the compatibility between these assumptions and the underlying hypothesis of molecular chaos.

We assume that cells are rod-shaped and when a given cell strikes a second cell, it instantly turns around its ``head" (the tip of the rod in the direction of motion) to match a second cell's orientation. Thus, in our model, the re-orientations are truly {\it nematic} and also {\it asymmetric}, in agreement with empirical observations of myxobacteria alignment, Fig \ref{fig:myxo}. For the asymptotic parameter regime, our model goes beyond the zero-order approximation used in the above-mentioned papers, by including the next-order correction. In fact, if one takes the values of model parameters from typical experiments with myxobacteria, one can see that the second-order correction term (in a non-dimensional equation) can be as large as 0.1 and, thus, is not negligible. We derive our main kinetic equation for the nematic alignment under these assumptions (section \ref{sec:1}, equation \eqref{eqn:kin3}).  In section \ref{sec:2}, we reduce this general kinetic equation to a system of numerically tractable PDEs by assuming only a finite number of orientations are present. The system turns out to be conservative and of the hyperbolic type. We then use a Lax-Friedrichs numerical algorithm, specifically designed to treat such systems, to obtain the numerical solutions.

Additionally, in section \ref{sec:2} we perform a comparative study of the numerical solutions of the PDEs with the agent-based simulations to determine the validity of the molecular chaos assumption. Our main finding here is that, unless an appreciable amount of noise is added to the model, the molecular chaos assumption does not hold. We show that even in simple cases the agent model exhibits cluster formation, with the rate of clustering increasing with the number of cells for the asymptotic regime considered.

As previously discussed, these results do not come as a complete surprise; instead, they validate the criticisms directed towards the Boltzmann framework when applied to scenarios involving self-propelled, rod-shaped cells engaged in asymmetric nematic alignment.

\section{Derivation of kinetic equation}

\label{sec:1}
Our goal is to derive a tractable kinetic description of a system of $N$ self-propelled rods in of length $l$ moving at constant speed $v$ in 2 dimensions, starting from a microscopic description of collisions.
Let $f(\bx,\theta,t)$ be the probability a rod with orientation $\theta$ has its head located at $\bx \in \mathbb{R}^2$ at time $t$. We will next derive a kinetic equation for $f$ by finding an expression for the change in $f$ over a single time step of size $\Delta t$. To this end, we make several assumptions about rod collisions. First, we assume that collisions are binary events, with the striking rod reorienting to match the struck rod's orientation modulo $\pi$. This results in the nematic alignment of rods. Motivated by \textit{M. xanthus} dynamics, we further assume that reorientation occurs through the head of the rod. Since we assume binary collisions, we will make use of the 2-particle distribution function $f_2(\bx,\theta,\bx_1,\theta_1,t)$ to help calculate collision events.

\begin{figure}[htbp]
\centering
    \includegraphics[clip, trim=0cm 2cm 0cm 2cm, width=0.5\textwidth]{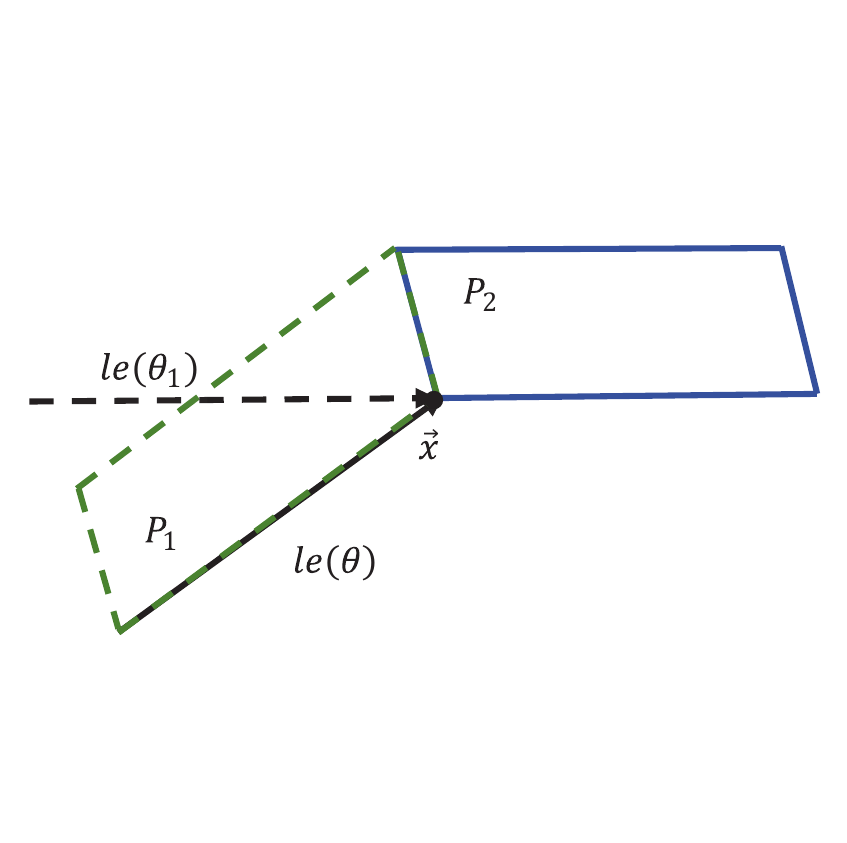}
    \caption{Geometry of the collision scheme considered in our model. $P_1$ is the set of heads of rods with orientation $\theta_1$ that will hit the side of the rod with orientation $\theta$ having its head at $\bx,$ at time $t,$ when it moves for $\Delta t$ units of time. Likewise, a rod with orientation $\theta$ at $\bx$ at time $t$ will hit the side of a rod with orientation $\theta_1$ having its head in $P_2$ at time $t.$ $P_1$ is formed by vectors $l{\bf e}(\theta)$ and $v\drmt({\bf e}(\theta_1)-{\bf e}(\theta)).$ $P_2$ is formed by vectors $l{\bf e}(\theta_1)$ and $v\drmt({\bf e}(\theta_1)-{\bf e}(\theta)).$    
    }
    \label{fig:2_1}
\end{figure}

Denote by $\Delta f$ the change of $f$ along the trajectory: 
\[\Delta f(\bx,\theta,t)=f(\bx+v{\bf e}(\theta)\drmt,\theta,t+\Delta t)-f(\bx,\theta,t),
\] 
with $N\Delta f$ representing the change in the number of rods with a given location and orientation.  In the absence of collisions, the change is zero. If a collision occurred it could result in the gain or loss of rods of orientation $\theta.$  The associated geometry of rods for gain and loss is determined by fixing a spatial location $\bx$ and considering two cases using two (non-interacting) rods with orientations $\theta$ and $\theta_1$ and heads at $\bx$. We then look at the sets of collisions in the next $dt$ time that will result in either a gain of a rod with orientation $\theta$ (due to a collision with a $\theta$-rod) or an equivalent loss (due to the $\theta$-rod colliding). This yields two regions $P_1(\bx,\theta,\theta_1)$ and $P_2(\bx,\theta,\theta_1)$ where another rod could be located to cause a collision (see Figure \ref{fig:2_1}).

We can now use this setup to derive the change $\Delta f.$
\begin{equation}
\begin{split}
\Delta f(\bx,\theta,t) =& 
-(N-1)\int_{-\pi}^{\pi}\int_{P_2(\bx,\theta,\theta_1)}f_{2}(\bx,\theta,\bx_1,\theta_1,t)\,d\bx_1d\theta_1\\
&+(N-1)\int_{\theta-\pi/2}^{\theta+\pi/2}\int_{P_1(\bx,\theta,\theta_1)}f_{2}(\bx,\theta_1,\bx_1,\theta,t)\,d\bx_1d\theta_1\\
&+(N-1)\int_{\theta-\pi/2}^{\theta+\pi/2}\int_{P_1(\bx,\theta+\pi,\theta_1)}f_2(\bx,\theta_1,\bx_1,\theta+\pi,t)\,d\bx_1d\theta_1\\
\end{split}
\end{equation}
To get a closed-form equation for $f$ we employ a commonly used assumption that 2-particle distribution $f_2(\bx,\theta,\bx_1,\theta_1,t)$ can be written as the product of the marginal distributions $f(\bx,\theta,t)$ and $f(\bx_1,\theta_1,t)$, i.e.
\begin{equation}\label{eqInd}
    f_2(\bx,\theta,\bx_1,\theta_1,t) {}={} f(\bx,\theta,t)f(\bx_1,\theta_1,t),
\end{equation}
for all pairs $(\bx,\theta)$ and $(\bx_1,\theta_1).$ This first-order moment closure is known as statistical independence, or molecular chaos, as it implies that the presence of one rod does not affect the probability of finding the other at the given position and orientation. Next, we write the integral terms using a Taylor expansion in $\bx_1$
\begin{equation}
f(\bx_1,\theta_1,t) = f(\bx,\theta_1,t) + \grad_\bx f(\bx,\theta_1,t)\cdot(\bx_1-\bx) + O(|\bx_1-\bx|^2),
\end{equation}
so that only an integral over $\theta_1$ remains. Note that $|\bx_1-\bx| = O(l)$ since two rods must be close to enable collision. A geometric computation shows that the regions $P_1$ and $P_2$ where collisions can take place have area $|P_1|=|P_2|=l v \drmt|\sin(\theta-\theta_1)|.$
Substituting the Taylor expansion into the expression for $\Delta_1 f$ and using the areas of $P_1$, $P_2$ with the integrals
\begin{equation}
\int_{P_2(\bx,\theta,\theta_1)} (\bx_1-\bx)\,d\bx_1{}={}\frac{l^2 v \drmt}{2}|\sin(\theta-\theta_1)| \, {\bf e}(\theta_1)+O(l^3v\drmt),
\end{equation}
and
\begin{equation}
\int_{P_1(\bx,\theta,\theta_1)} (\bx_1-\bx)\,d\bx_1{}={}\frac{l^2 v \drmt}{2}|\sin(\theta-\theta_1)| \, {\bf e}(\theta)+O(l^3v\drmt),
\end{equation}
we obtain
\begin{equation}
\begin{split}
&    \Delta f(\bx,\theta,t) =(N-1)l v \drmt\int\limits_{\theta-\pi/2}^{\theta+\pi/2} |\sin(\theta-\theta_1)|f(\bx,\theta+\pi,t)f(\bx,\theta_1,t)\,d\theta_1\\
&    {}-{}
    (N-1)l v \drmt\int\limits_{\theta-\pi/2}^{\theta+\pi/2} |\sin(\theta-\theta_1)|f(\bx,\theta,t)f(\bx,\theta_1+\pi,t)\,d\theta_1\\
&    {}+{}\frac{(N-1)l^2 v \drmt}{2}\int\limits_{\theta-\pi/2}^{\theta+\pi/2}|\sin(\theta-\theta_1)|f(\bx,\theta_1,t)\left({\bf e}(\theta)\cdot\grad_\bx f(\bx,\theta,t)+{\bf e}(\theta+\pi)\cdot\grad_\bx f(\bx,\theta+\pi,t)\right)\,d\theta_1 \\
&    {}-{}\frac{(N-1)l^2 v \drmt}{2}\int\limits_{\theta-\pi/2}^{\theta+\pi/2}|\sin(\theta-\theta_1)| f(\bx,\theta,t)\left({\bf e}(\theta_1)\cdot\grad_\bx f(\bx,\theta_1,t)+{\bf e}(\theta_1+\pi)\cdot\grad_\bx f(\bx,\theta_1+\pi,t)\right)\,d\theta_1\\
&     {}+{}O(Nl^3v\drmt\max|f|\max|\grad_\bx^2f|){}+{}O(Nl^2\drmt^2).
\end{split}
\end{equation}
Dividing 
the equation above
by $\Delta t$ and letting $\Delta t\to 0,$ we obtain the kinetic equation
\begin{equation}\label{eqn:kin1}
\partial_t f+v{\bf e}(\theta)\cdot\grad_\bx f =
(N-1)lv Q_0 + \frac{(N-1)l^2v}{2}Q_1 + O\left(Nl^3v\max|f|\max|\grad_\bx^2f|\right),
\end{equation}
where
\begin{equation}
Q_0 {}={}\int\limits_{\theta-\pi/2}^{\theta+\pi/2} |\sin(\theta-\theta_1)|\left(f(\bx,\theta+\pi,t)f(\bx,\theta_1,t)-f(\bx,\theta,t)f(\bx,\theta_1+\pi,t)\right)\,d\theta_1,
\end{equation}
and

\begin{equation}
\begin{split}
Q_1{}&={}\int\limits_{\theta-\pi/2}^{\theta+\pi/2}|\sin(\theta-\theta_1)|f(\bx,\theta_1,t){\bf e}(\theta)\cdot\grad_\bx\left(f(\bx,\theta,t)-f(\bx,\theta+\pi,t)\right)\,d\theta_1\\
&-\int\limits_{\theta-\pi/2}^{\theta+\pi/2}|\sin(\theta-\theta_1)|f(\bx,\theta,t){\bf e}(\theta_1)\cdot\grad_\bx (f(\bx,\theta_1,t)-f(\bx,\theta_1+\pi))\,d\theta_1.
\end{split}
\end{equation}
In deriving $Q_1$ term we used the identity ${\bf e}(\theta+\pi)=-{\bf e}(\theta).$

We nondimensionalize equation \eqref{eqn:kin1} by setting $\tau{}={}L/v$ and rescaling the variables using 
\begin{equation}
\hat{\bx}=\bx/L,\quad \hat{t}= t/\tau,\quad \hat{f}(\hat{\bx},\theta,\hat{t})=L^{2} f(\hat{\bx}L,\theta,\hat{t}\tau).
\end{equation}
Dropping hats, the resulting scaled kinetic equation is
\begin{equation}
\partial_t f+{\bf e}(\theta)\cdot\grad_\bx f =
\frac{(N-1)l}{L} Q_0 + \frac{(N-1)l^2}{2L^2}Q_1 + O\left(\frac{Nl^3}{L^3}\right),
\end{equation}
where we assumed that the variations in the kinetic density $f$ are bounded, that is $\max|f|\max|\grad^2_\bx f|\leq C.$ For a typical experiment with $N=1000$ \textit{M. xanthus} bacteria ($l=5\mu m$) on a domain of size $L=10^3\mu m,$  parameter $\frac{Nl}{L}=5$ and $\frac{Nl^2}{L^2}=0.025,$ and $\frac{Nl^3}{L^3}=0.000125.$ Thus, when $C$ is of order 1, a reasonable approximation is keeping only the first two terms on the right-hand side. We set $\kappa = \frac{N l^2}{2 L^2}.$ The parameter $\kappa$ is related to the mean free path, $d,$ and rod length $l$ by
\[
\kappa\leq \frac{l}{2d}.
\]
See \eqref{eq:kappa_bound} of the Appendix for details. The binary collision assumption fails if $d$ is of order $l$ or smaller, meaning that $\kappa$ should be small. In the numerical simulations in the subsequent sections, we will restrict $\kappa<0.5.$
The resulting kinetic equation becomes
\begin{equation}\label{eqn:kin3}
\partial_t f+{\bf e}(\theta)\cdot\grad_\bx f =
\frac{2\kappa L}{l}Q_0(f)+\kappa Q_1(f).
\end{equation}

In the model of alignment that we consider in this paper, a rod after an interaction assumes the orientation of the rod it collides with. The set of orientations $\theta_i,$ $\theta_i+\pi,$ 
$i=1,\ldots,k$, if
present initially, will be preserved by the dynamics of transport and collision. In such situations, the kinetic density $f$ is determined by the set of $2k$ densities of the corresponding orientations, and equation \eqref{eqn:kin3} can be written as a system of $2k$ partial differential equations for the orientation densities. An example of such a system is considered in the following sections.



\section{Numerical solutions and results from agent-based simulations for case study}
\label{sec:2}
In the absence of correlations in the underlying system, kinetic equations derived from Boltzmann's hypothesis should accurately model the system's evolution in time. However, the presence of correlations can cause discrepancies to appear. Here we will test the accuracy of our kinetic equation by comparing the behavior of numerical solutions to agent-based simulations of the underlying system. We will use a test case to demonstrate that even for simple setups, the microscopic rules of collision lead to a buildup of orientational correlations, causing the correlation-free model to underestimate the effects of rod-rod interactions. While this test case does not capture the complexity present in many physical systems, it is enough to demonstrate the need to account for correlations between rods when modeling active matter systems such as bacterial communities.

\subsection{A model with two orientations}
\label{sec:two_conservative}
Consider equation \eqref{eqn:kin3} for distributions with $n$ fixed orientations $f(\bx,\theta,t) = \sum_{i=1}^n \rho_i(\bx,t)\delta(\theta - \theta_i)$, $\bx = [x, y]^T$. In general, this will result in a system of $n$ non-conservative PDEs in two spatial dimensions, making accurate numerical solutions over long periods of time hard to obtain. To illustrate our point about growth of correlations, we will focus on a reduced case with two orientation angles $\theta_1$ and $\theta_2$ where $|\theta_2-\theta_1|\leq \pi/2.$ Then there is no nematic alignment and thus the term $Q_0(f)$ vanishes. Additionally, as we will show, the system of two PDEs can be written in conservation form under certain assumptions about the initial density distributions.
Substituting $f(\bx,\theta,t) = \rho_1(\bx,t)\delta(\theta - \theta_1) + \rho_2(\bx,t)\delta(\theta - \theta_2)$ into \ref{eqn:kin3} yields a system for ${\bf \rho} = (\rho_1,\rho_2)$ of the form
\begin{equation}
    \label{Model3:2angles}
    \partial_t{\bf \rho} + A({\bf \rho})\partial_x{\bf \rho} + B({\bf \rho})\partial_y{\bf \rho} {}={} 0
\end{equation}
where the matrices $A$ and $B$ are given by
\[
A{}={}
\left[
\begin{array}{rr}
\cos(\theta_1)+\kappa\cos(\theta_2)\rho_2 &\quad -\kappa\cos(\theta_1)\rho_1 \\
-\kappa\cos(\theta_2)\rho_2 &\quad \cos(\theta_2)+\kappa\cos(\theta_1)\rho_1
\end{array}
\right],
\]
\[
B{}={}
\left[
\begin{array}{rr}
\sin(\theta_1)+\kappa\sin(\theta_2)\rho_2 &\quad -\kappa\sin(\theta_1)\rho_1 \\
-\kappa\sin(\theta_2)\rho_2 &\quad \sin(\theta_2)+\kappa\sin(\theta_1)\rho_1
\end{array}
\right].
\]

To simplify these kinetic equations to a form easily analyzed, we make two assumptions: that all densities vary only in the spatial direction $x$ so the problem is effectively 1D with $\rho(\bx,t) = \rho(x,t)$, and that $\theta_1=\frac{\pi}{4},$ $\theta_2=\frac{3\pi}{4}$ so the system can be put into conservation form. In this setting, the system (\ref{Model3:2angles}) is simply
\begin{equation}\label{Model4:2angles_sim}
    \begin{split}
    \partial_t\rho_1 + \frac{\sqrt{2}}{2}\partial_x\left(\rho_1\left[1-\kappa\rho_2\right]\right)  {}&={}0\\
    \partial_t\rho_2- \frac{\sqrt{2}}{2}\partial_x\left(\rho_2\left[1-\kappa\rho_1\right]\right)  {}&={}0.
    \end{split}
\end{equation}
We restrict our analysis to the regime $0 < \kappa < 0.5$ to enforce the mean free path of rods is greater than the rod length $l$ (see Appendix \ref{append:mfp}). We will consider a test case with two opposing waves of rods over a constant background density. Each wave is a Gaussian moving over a uniform background density of rods. The initial conditions for $\rho_1$ and $\rho_2$ are of the form
\begin{equation}\label{Model4:ics}
    \rho_i(x,0) {}={} a_i+\frac{b_i}{\sqrt{2\pi}\sigma_i} e^{-(x-c_i)^2/(2\sigma_i^2)}
\end{equation}
with the constraints $\int_0^1\int_0^1 \rho_1(x,0) + \rho_2(x,0) dxdy = 1$ from conservation and an enforced constraint $a_i/b_i = r_i$ for the ratio of background cells to the cells forming the waves. The final parameters $c_i$ and $\sigma_i$ then control the location and width of the waves. Additionally, the maximum amplitude $a_i+b_i/(\sqrt{2\pi}\sigma_i)$ set below the bound established in Appendix \ref{append:hyper}. As time evolves, the linear factor $1-\kappa\rho_i$ inside the spatial derivative will decrease both the left-and right-moving waves' speeds when they interact. Finally, we set $c_1 = 0.25$, $c_2 = 0.75$, $\sigma_1 = \sigma_2 = 0.0625$, and $r_i = 7$ so the two waves' peaks will move towards each other and meet at $x = 0.5$.  We numerically solve the system \eqref{Model4:2angles_sim} using the method described in Appendix \ref{append:num}.

\subsection{Agent-based model setup}

For the agent-based model, we set the domain size to $L = 400$, the velocity to $v = \sqrt{2}$. Each rod $i = 1, \dots, N$ followed the equations of motion
\begin{equation}\label{Model:agents}
    \begin{split}
    \frac{d\bx_i}{dt} {}&={} v[\cos(\theta_i), \; \sin(\theta_i)]^T \\
    \frac{d\theta_i}{dt}  {}&={} (\theta_k-\theta_i)\delta(t-T_{i,k}).
    \end{split}
\end{equation}
where $\delta(t)$ is the Delta distribution and $T_{i,k}$ denotes the times when the $i$th rod collides with another rod $k$ and reorients. The initial location of each rod was determined through random sampling of the initial density profiles $\rho_i(x,0)$. Collisions between rods were determined by implementing the schematic in Figure \ref{fig:2_1}. If a rod would collide with multiple rods in a single time step, it was assumed that it collided with the first rod in its path to ensure collisions are binary. We restricted the mean-free path of rods to be greater than the rod length $l$ based on the physical timescale of reorientation $l/v$ (see Appendix \ref{append:mfp}). Therefore, we took time steps of $\Delta t = l/v$. This discretization also has the advantage of reducing computational time.

The simulations were deterministic after initial rod positions are sampled, so we combined 1000 simulations to smooth fluctuations and recover the mean behavior. Since the solutions of system \eqref{Model4:2angles_sim} vary spatially only in the $x$-direction, we compared them to agent-based simulations by looking only at the $x$-coordinates of simulated rods. We then constructed a 1D kernel density estimate (KDE) using these values. The resulting density profiles were then scaled to probability densities on $x \in [0, \, 1]$ so they could be compared to the density profiles obtained from numerically solving the kinetic equations.

\subsection{Discrepancy between agent-based model and kinetic equations due to cluster formation}

\begin{figure}[htbp]
\centering
    \includegraphics[clip, trim=0cm 2.5cm 0cm 2.5cm, width=\textwidth]{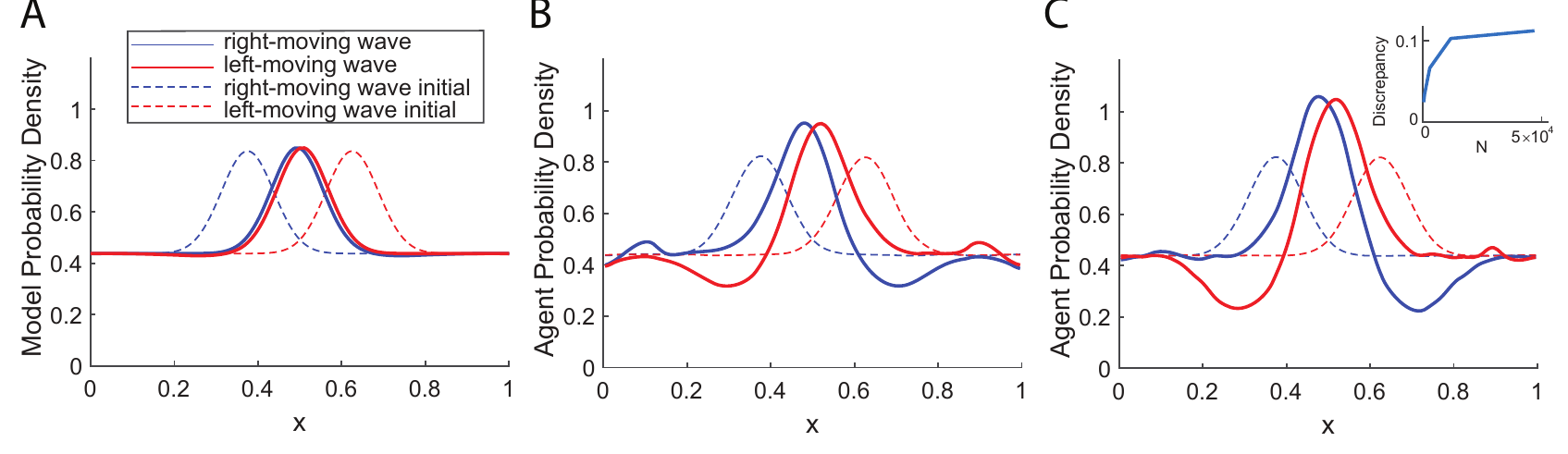}
    \caption{A) Numerical solution of the kinetic equations for $\kappa = 0.1$. B) Density profile for agent-based simulations for $\kappa = 0.1$ with $N = 2856$ rods. C) Density profile for agent-based simulations for $\kappa = 0.1$ with $N = 45700$ rods. (Inset) The discrepancy at the final time between the kinetic model density profile and the density profile from agent simulations for various numbers of rods. The results indicate that the agent simulations show much greater changes in the density profile and do not converge to the kinetic model in the limit $N \to \infty$, $l \to 0$.}
    \label{fig:4}
\end{figure}

Our agent-based simulations deviate from the numerical solutions of the kinetic equations, with the latter exhibiting a far less pronounced change in the density profile shape (Figure \ref{fig:4} A\&B). The peaks of the density profiles for both orientations slow down slightly as they approach each other, but the slowdown and the increase in the peak density are much greater in the agent framework. If this discrepancy is due to not being near the limit $N \to \infty$, $l \to 0$ with $\kappa$ held constant, increasing the number of simulated rods while keeping $\kappa \propto Nl^2$ fixed should theoretically improve the agreement. However, we see that this is not the case (Figure \ref{fig:4}C). The simulations using higher numbers of rods showed greater discrepancies. This suggests the disagreement is due to one or more flaws in the assumptions used to derive the kinetic equations.

The key assumption used in deriving the Boltzmann-type equation was the statistical independence of the joint probability distribution \eqref{eqInd}. This allowed us to express the probability that a pair of rods would have a spatial configuration leading to a collision in terms of the probability of each rod individually occupying the corresponding spatial region. This assumption is invalid if there are correlations between rods. Such correlations appear in clusters of aligned rods since they have a high chance of having similar orientations due to collisions. Our collision scheme results in both rods possessing the same orientation, so cluster formation is possible and would be a violation of our assumption of statistical independence.

To measure the clustering present in our agent simulations, we used a standard algorithm based on Euclidean distance to group rods. The minimum distance between the heads of rods was picked to be the length of a rod $l$. We then measured the proportion of rods in clusters of size 4 or more (Figure \ref{fig:5}A). As expected, clusters grew over time in our agent simulations, with the rate of growth increasing with $\kappa$ for fixed $N$. This increase in $\kappa$ corresponds to a greater rod length, increasing the chance of collisions between rods.

\subsection{Quantifying loss of statistical independence}
Since cluster formation is linked to a loss of statistical independence \cite{hanke_understanding_2013,thuroff_critical_2013,ihle_towards_2014}, we next quantified the extent of this loss. Observing a rod with orientation $\theta_1$ would decrease the probability of a nearby rod having orientation $\theta_2$. This is typically captured using two-particle or higher correlation functions. If the system is spatially homogeneous or can be approximated as such, then there are several ingenious ways to approach calculating these quantities \cite{jeggle20,kursten20,kursten21}. This is commonly used to analyze systems near the onset of polar or nematic order. In our case, we are far from spatial homogeneity, so we use a different approach.

In our work, we directly calculated metrics from agent simulation data related to the quantity $f_2(x_1,\theta_1,x_2,\theta_2,t) - f(x_1,\theta_1,t)f(x_2,\theta_2,t)$ to see if a loss of independence is present. We first divided our domain $\Omega = [0, \, L] \times [0, \, L]$ into $2^{2m}$ square subregions $\Omega_{ij}$, $i,\, j = 1,\dots,2^{m}$ with side length $L/2^{m}$. Since subregions should be large enough to contain small clusters of rods, we set $m = 5$ so that $L/2^{m} = 12.5$. We pooled all rods appearing in $\Omega_{ij}$ across 1000 simulations with different initial agent locations $\bx_i(0)$, then we calculated coarse-grained approximations of the joint and marginal distributions conditioned on rods being in $\Omega_{ij}$
\begin{equation}\label{eqn:conditional}
    \begin{split}
    f_2^{ij}(\theta_1,\theta_2,t) &= \frac{\mbox{\# pairs of rods with orientations $(\theta_1,\theta_2)$ in box}\; ij}{\mbox{\# pairs of rods in box}\; ij} \\
    f^{ij}(\theta_1,t) &= \frac{\mbox{\#rods with orientation $\theta_1$ in box}\; ij}{\mbox{\#rods in box}\; ij} \\
    f^{ij}(\theta_2,t) &= \frac{\mbox{\# rods with orientation $\theta_2$ in box}\; ij}{\mbox{\# rods in box}\; ij} 
    \end{split}.
\end{equation}
To measure the loss of statistical independence based on orientations in $\Omega_{ij}$ at time $t$, we chose Pearson's correlation coefficient calculated from the estimates of the conditional joint and marginal
\begin{equation}\label{Model4:correlation}
r_{\theta_1\theta_2}^{ij}(t) = \frac{\left( \sum_{m,n = 1,2} \theta_n \theta_m f_2^{ij}(\theta_n,\theta_m,t) \right) - M^2}{\sum_{n=1,2} (\theta_n - M)^2f^{ij}(\theta_n,t)}
\end{equation}
where $M$ the mean orientation calculated from $f^{ij}(\theta_1,t)$ and $f^{ij}(\theta_1,t)$.
Averaging over all subregions then gives us an average correlation $r_{\theta_1\theta_2}$.

\begin{figure}[htbp]
\centering
    \includegraphics[clip, width=\textwidth]{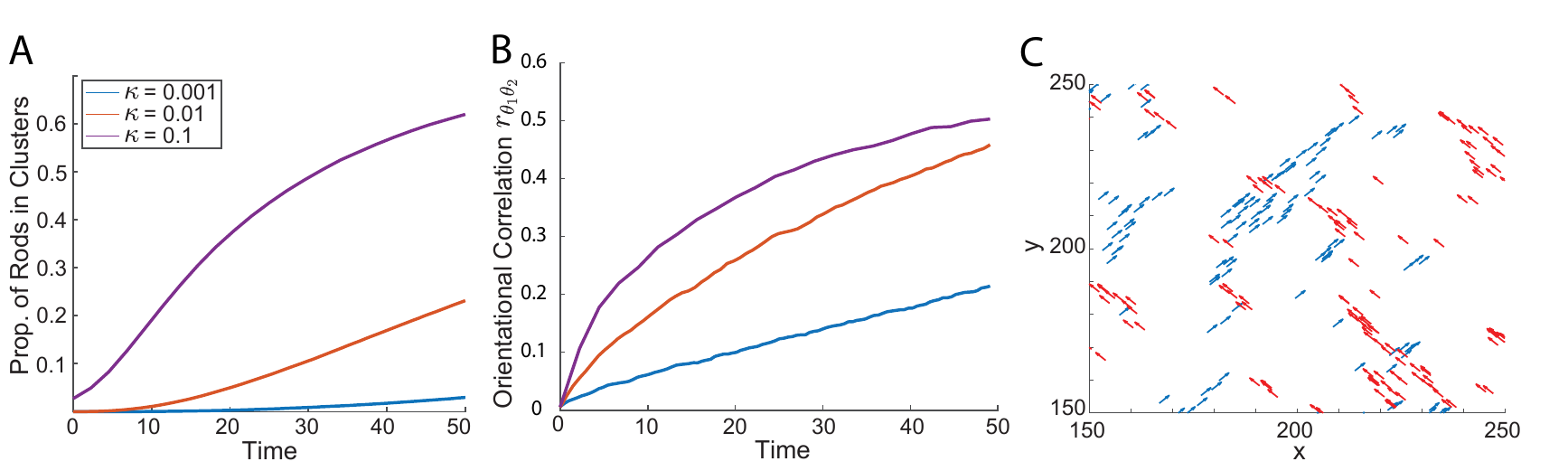}
    \caption{A) The percentage of rods in clusters with more than 4 rods over time. This quantity grows over time as rods collide and co-orient. B) The loss of statistical independence measured via a local orientation correlation coefficient $r_{\theta_1\theta_2}(t)$. Here the correlations growing over time indicate that it becomes rarer to see nearby rods having different orientations, e.g. cluster formation. C) Snapshots of an agent simulation at the end of a simulation. The snapshot is taken at the center of the domain where the two waves interact.}
    \label{fig:5}
\end{figure}

As shown in Figure \ref{fig:5}B, the metric $r_{\theta_1\theta_2}(t)$ shows similar trends over time to the cluster formation. The loss of statistical independence increases over time, with higher values of $\kappa$ showing a greater loss for fixed $N$.This reflects the fact that the chance of finding a pair of rods close together with different orientations is lowered from what is expected if statistical independence holds.

\begin{figure}[htbp]
\centering
    \includegraphics[clip, width=\textwidth]{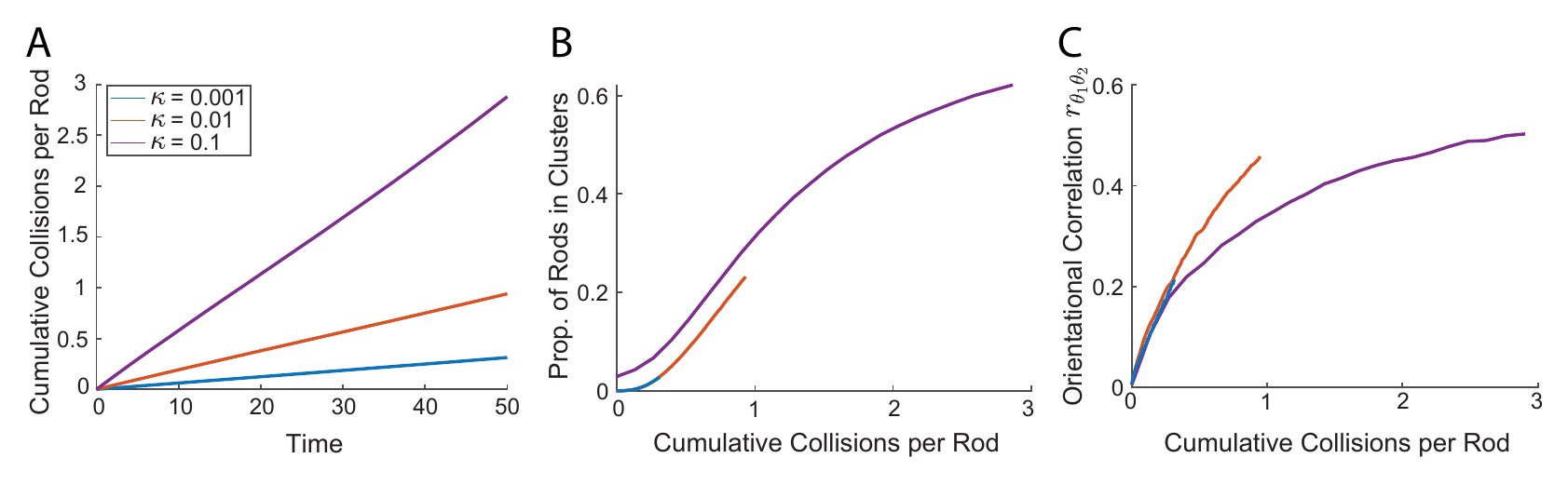}
    \caption{A) The total number of collisions experienced on average per rod over time for $\kappa = 0.001$ (blue), $\kappa = 0.01$ (orange), and $\kappa = 0.1$ (purple). B) The fraction of rods in clusters with more than 4 rods versus the mean number of collisions experienced per rod by time $t$. This is plotted for $\kappa = 0.001$ (blue), $\kappa = 0.01$ (orange), and $\kappa = 0.1$ (purple) C) The orientational correlation coefficient $r_{\theta_1\theta_2}(t)$  between nearby pairs of rods versus the mean number of collisions experienced per rod by time $t$. Both figures show some data collapse when compared to Figure \ref{fig:4}, especially for low mean rod collisions per rod.}
    \label{fig:6}
\end{figure}

Since cluster formation is the result of co-alignment between rods upon collision, it is natural to see if the increase in cluster growth with higher values of $\kappa$ is due to a greater number of collisions. Calculating the mean number of collisions per rod by time $t$ shows that the number does increase as $\kappa$ increases (Figure \ref{fig:6}A). Furthermore, there is strong evidence of data collapse when plotting both the proportion of rods in clusters (Figure \ref{fig:6}B) and the loss of statistical independence versus the cumulative number of collisions per rod (Figure \ref{fig:6}C). There is a slight difference in the rescaled curve for $\kappa = 0.1$ compared to the other two curves. This is the result of some rods starting in clusters at $t = 0$ at higher densities. The reasons for the deviation in the rescaled $r_{\theta_1\theta_2}$ cure for $\kappa = 0.1$ is less clear but could be the result of cluster-cluster interactions once most rods reside in such clusters. The rescaling is based on rod-rod collisions, so cluster-cluster interactions might not fully be accounted for.

\subsection{Improving agreement between agent simulations and kinetic equations by inhibiting cluster formation}

While the results of our simulations have indicated close links between cluster formation, loss of independence, and the discrepancy between the equations and agent simulations, they have not shown a strict cause and effect. To illustrate that clustering is the main cause of the discrepancy, we introduce diffusion in $y$-direction into our agent simulations. The addition of this noise will cause rods forming a cluster to slowly drift apart at a rate dependent on the noise strength. This addition will not impact the test case we considered for the kinetic model. Since initial conditions are constant in the $y$-direction, the system given in \ref{Model4:2angles_sim} will remain unchanged by the vertical diffusion. 

\begin{figure}[!htbp]
\centering
    \includegraphics[clip, trim=0cm 1cm 0cm 1cm, width=0.8\textwidth]{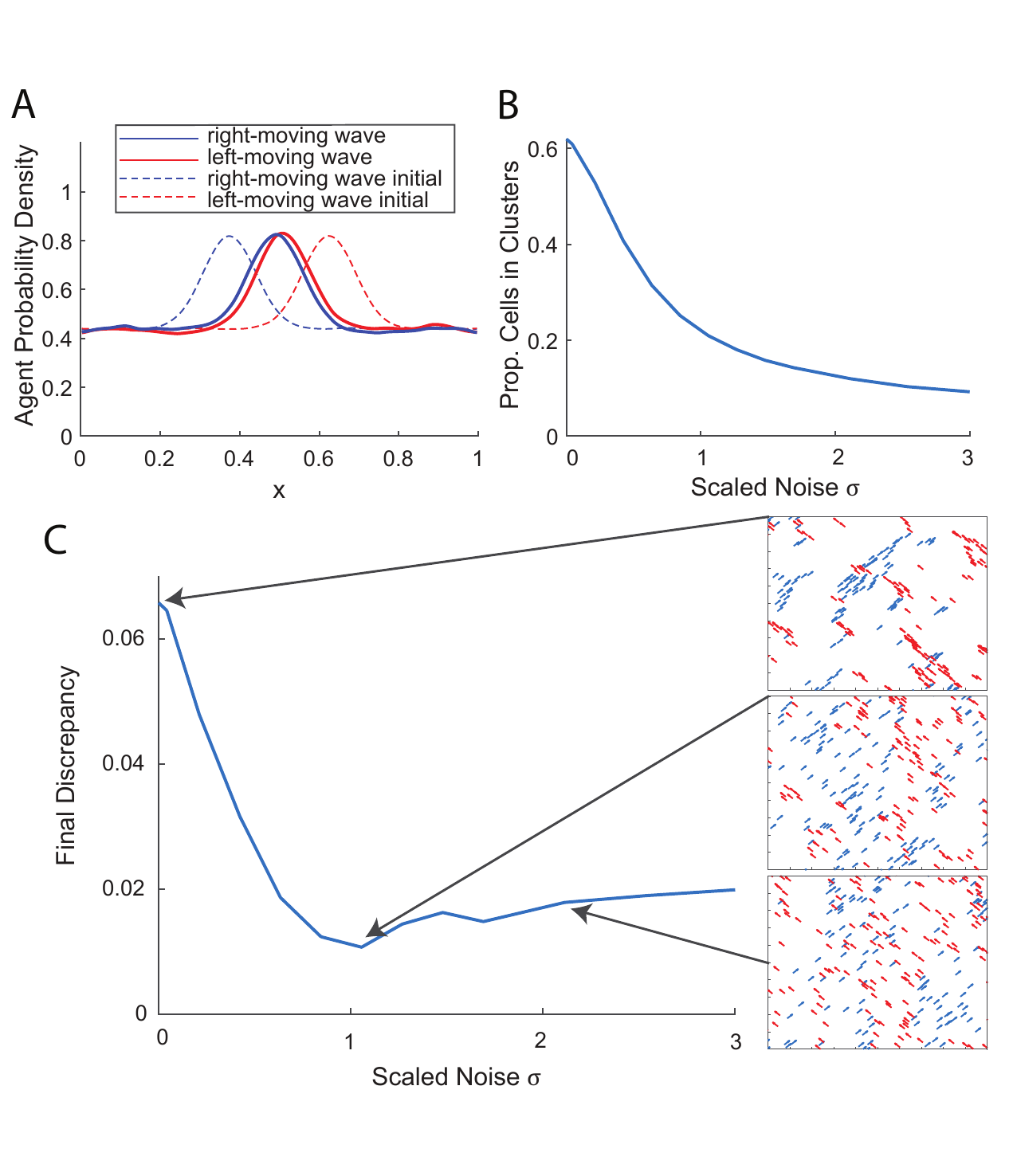}
    \caption{A) Density profile for agent-based simulations with $y$-directional scaled noise of $1.06$ for $\kappa = 0.1$ with $N = 2856$ rods. A scaled noise level of 1 corresponds to an agent rod's vertical movement over time $\Delta t = l/v$ being distributed normally as $y(t+dt)-y(t) \sim N(l \sin(\theta_i), l \sin(\theta_i)) = N(l/\sqrt{2}, l /\sqrt{2})$ in between collisions, or equivalently to a vertical diffusion coefficient of $l^2/4$.   B) Percentage of rods in clusters with more than 4 rods versus noise in $y$-direction. Increasing the noise in the vertical direction decreases the proportion of rods in larger clusters C) Discrepancy between the kinetic model and the agent-based simulations for different levels of white noise in the $y$-direction. The discrepancy initially drops sharply but then increases slightly once the strength of the noise increases past $1$, the length of a rod projected in the y-direction. Snapshots of agent simulations near noise levels of $0$, $1$, and $2$ are shown on the right. The snapshots are taken at the final time in the center of the domain ($[150,\, 250]\times[150,\, 250]$) where the two waves interact.}
    \label{fig:7}
\end{figure}

We implemented the same sets of agent simulations as before ($N = 2856$, $\kappa = 0.1$) with the addition of different levels of noise $\sigma$. Agent rods step in the $y$-direction a random distance drawn from a normal distribution $N(0,\sigma)$ every $\Delta t = l/v$. Since agent simulations are run with unscaled variables, we nondimensionalize the noise by using the scaling $\sigma \rightarrow \sigma/(l/\sqrt{2})$ in order to compare the strength of the noise to the rod length $l$. The results of these simulations show that the addition of noise in the $y$-direction improves agreement between the agent simulations and the kinetic equations, with the former now resembling the latter for a sufficient level of noise (Figure \ref{fig:7}A). Increasing the noise strength from zero reduces the measured discrepancy between the kinetic and agent density profiles (Figure \ref{fig:7}B) and decreases the proportion of rods in clusters (Figure \ref{fig:7}C). However, the discrepancy is reduced only up to a point. Once the strength of the scaled noise exceeds roughly $1$ (corresponding to the standard deviation of the normal distribution equaling the projection of a rod in the y-direction), the discrepancy increases slightly before plateauing. Why this is the case is unclear; however, the reduction in clustering slows down at around the same level of noise. It is possible that there are some aspects of rod correlations that the added diffusion does not affect. For example, our collision scheme results in both rods having similar x-coordinates. The addition of vertical diffusion does not change this either. Therefore, successive collisions result in more rods sharing similar x-coordinates, and potentially forming vertical bands of rods, regardless of vertical diffusion.

\newpage 

\section{Discussion}
In this paper, we developed a kinetic model for the alignment of self-propelled hard rods where collisions result in asymmetric alignment. Using this model, we showed Boltzmann formalism severely underestimates the change in rod density profiles when two opposing waves of rods interact. We explicitly measured the loss of statistical independence that invalidates the classic assumption of molecular chaos Boltzmann-type equations rely on. Such a loss corresponds to the formation of rod clusters due to alignment from binary rod collisions. Our results mirror those in other studies \cite{hanke_understanding_2013,thuroff_critical_2013}, however, we have built upon these works by showing that mechanisms that destroy or inhibit cluster formation help restore agreement between the kinetic model and agent-based implementations of the microscopic alignment rules. As this discrepancy occurs even in the simple setup we consider, our work highlights the need to extend current methodologies beyond Boltzmann-type kinetic equations in order to accurately capture the properties of active matter in biological systems.


Boltzmann's hypothesis can be justified when the mean free path is large compared to the range of local interactions, however, this is rarely satisfied at realistic densities when ordered motion is established \cite{pismen_active_2021}. Additionally, several studies have indicated that rapid cluster formation can lead to a strong violation of the molecular chaos assumption needed in the Boltzmann approach \cite{hanke_understanding_2013,thuroff_critical_2013}. While gradual alignment results in a better match, in principle any collision rule between particles resulting in alignment can cause correlations to appear. Such collisions are prevalent in collective dynamics at the cellular level due to a low Reynolds number, where cells must actively exert energy to maintain their motion. This is in contrast to Boltzmann gas dynamics, where collisions are assumed to be non-elastic and conserve momentum, resulting in particles simply bouncing off each other. In addition to violations of Boltzmann's hypothesis, the assumption of binary interactions used in such models is not enough to reproduce the observed dynamics in some biological systems \cite{suzuki_polar_2015}. Binary interactions can simply be too weak to produce the alignment seen experimentally, even when corrections are made to account phenomenologically for correlations. These breakdowns of Boltzmann-type models suggest they are an overly-simplistic approach to modeling the emergence or stability of collective alignment. The emergence of local order almost by definition involves the breakdown of statistical independence and the growth of correlations as agents align and start moving together, violating Boltzmann's hypothesis.

There are several ways to extend kinetic models to incorporate orientational correlations. Perhaps the simplest is to explicitly extend the kinetic model to explicitly include equations for clusters of various sizes. Gain and loss terms for these additional equations are then the natural result of collisions among clusters and rods \cite{peruani10,weber_role_2013}. Peruani et al (2010) developed a framework where cluster size was explicitly tracked in a hierarchy of equations, while Weber et al (2013) \cite{weber_role_2013} considered a simplified framework with two reaction equations for clusters and single cells. Another set of approaches for incorporating correlations involves direct modifications to the moment closure method. A simple example is replacing the joint distribution $f_2(x_1,\theta_1,x_2,\theta_2,t)$ with $\chi(\theta_2-\theta_1)f(x_1,\theta_1,t)f(x_2,\theta_2,t)$, where $\chi(\theta_2-\theta_1)$ is a phenomenological term accounting for correlations between different angles \cite{suzuki_polar_2015}. More sophisticated methods involve higher-order moment closures of the BBGKY hierarchy \cite{Bogoliubov1946, balagam14b, Grad1949}) using the so-called cluster expansion that explicitly incorporates the evolution of the joint distribution $f_2$ \cite{CercignaniBook}. Such an approach was used by Chou and Ihle for Vicsek-style models to extend beyond mean-field theory \cite{chou_active_2015}. With the rapid advancement of biological studies in the last decades, developing and applying new analytical models to understand active matter in biology is of crucial importance. Creating a tractable class of models that can capture correlations or non-binary interactions would provide a cornerstone for this young field.

\section{Acknowledgments}
Research is supported by National Science Foundation Division of Mathematical Sciences awards 1903270 (to M.P. and I.T.) and 1903275 (to OAI and PM).
We also would like to thank prof. D. Kuzmin (TU Dortmund) for helpful discussions on the numerical methods for the kinetic model presented in this paper.

\newpage


\begin{appendix}

\section{Mean free path}\label{append:mfp}

\begin{figure}[!htbp]
\centering
\includegraphics[clip, trim=0cm 5.5cm 0cm 5.5cm, width=\textwidth]{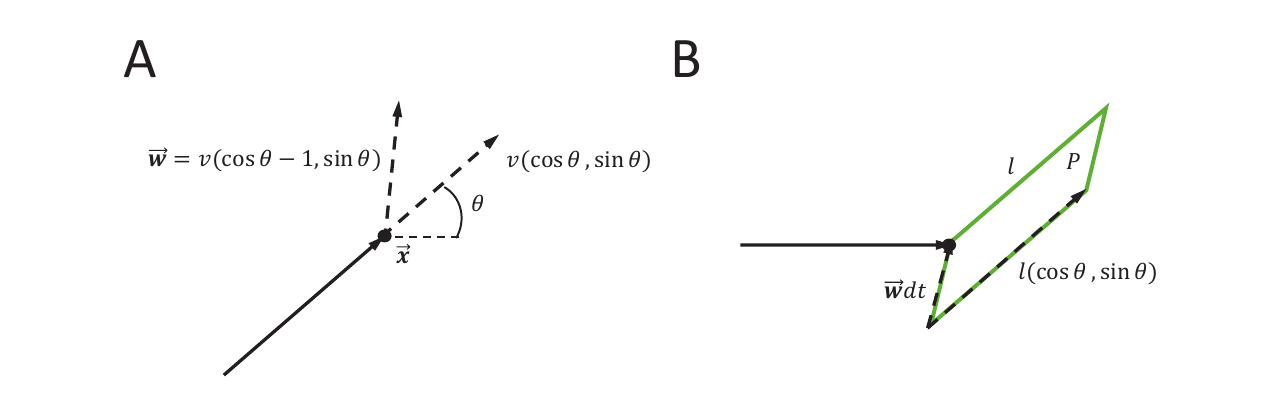}

\caption{Vector geometry used in the estimation of the MFP. A) rod at $x$ where ${\bf e}(\theta)=(\cos\theta,\sin\theta)$ moves with velocity $v {\bf e}(\theta).$ In a reference coordinates of a horizontally moving rod with velocity $(v,0),$ the rod velocity is ${\bf w}.$ B) a rod moving horizontally to the right, during time interval $[t^*,t^*+\Delta t]$ can collide with a rod which has orientation ${\bf e}(\theta)$ whose center is located in the parallelogram P, formed by vectors ${\bf w}\drmt$ and $l{\bf e}(\theta).$ The area of the parallelogram equals $|\sin\theta|v l \drmt.$
\label{mfp}}
\end{figure}

The mean free path (MFP) is defined as the average distance that a rod moves between collisions. This distance can be computed as the product of the speed $v$ and the time $\tau$ a rod moves between re-orientations. To estimate $\tau,$ we first compute the number of rods with orientation ${\bf e}(\theta)=(\cos\theta,\sin\theta)$ that 
will collide with a given rod (rod 1) in the time $[t^*, t^*+\Delta t]$, were $t^*$ marks the last change in rod 1's orientation. Without loss of generality, we consider rod 1 as moving horizontally to the right, as in Figure \ref{mfp}. Suppose that a coordinate system is chosen so that the rod is at rest. Then the velocity of rods with orientation $\theta$ is described by the vector ${\bf w}{}={}v(\cos\theta-1,\sin\theta).$ The heads of the rods that rod 1 can collide with are located in a parallelogram $P$ formed by vectors ${\bf w}\drmt$ and $l {\bf e}(\theta).$ The area of this parallelogram equals $|\sin\theta|vl\drmt.$ Using that $Nf(x,\theta,t^*)$ is the local number of rods with a given orientation, we estimate the number of collisions with $\theta$-oriented rods in time $dt$ as $vl N f(x,\theta,t^*)|\sin\theta|\drmt,$ which is maximized at $\theta=\pi/2.$ 
Thus, the number of collisions over all orientations in $\Delta t$ time can be estimated as
\[
v l N \drmt\int f(x,\theta,t^*)\,d\theta{}\leq{} n(x) v l \drmt,
\]
where $n(x,t^*)$ is the local rod number density.
From this, we obtain an upper bound for the frequency of collisions as $n(x,t^*)vl$. The time between collisions and the MFP are then bounded by
\[
\tau \geq \frac{1}{n(x,t^*)vl},\quad 
d \geq \frac{1}{n(x,t^*)l}.
\]
Note that both $\tau$ and the MFP $d$ are local quantities depending on $x.$

For the model we consider in this paper, we restrict the MFP to $d\geq l$. The rationale behind this is physics-based. When two rods collide, there is a characteristic time scale for reorientation to occur given by $\tau_\theta = l/v$ \cite{balagam15b}. Therefore it makes sense to restrict the mean time between collisions $\tau$ to the regime $\tau \geq \tau_\theta = l/v$. This naturally yields $d \geq l$.

Our assumption that $d\geq l$ puts a restriction on the local density of the form $n(x,t^*)l^2\leq1,$ and subsequently on the parameter $\kappa$ of the form
\begin{equation}
\label{eq:kappa_bound}
\kappa = \frac{Nl^2}{2L^2}\leq \frac{1}{2}l^2\max_x\,n(x,t^*)\leq\frac{1}{2}.
\end{equation}
Here we use that $N/L^2$ is a lower bound on the maximum of $n(x,t^*).$ Note that we cannot have $\kappa = 0$ without the local density being identically 0.

\section{Domain of hyperbolicity of system of equations\eqref{Model4:2angles_sim}}\label{append:hyper}
Here we will determine conditions under which system \eqref{Model4:2angles_sim} is hyperbolic. The system is given by
\begin{equation}
    \begin{split}
    \partial_t\rho_1 + \frac{\sqrt{2}}{2}\partial_x\left[\rho_1(1-\kappa\rho_2)\right]  {}&={}0\\
    \partial_t\rho_2- \frac{\sqrt{2}}{2}\partial_x\left[\rho_2(1-\kappa\rho_1)\right]  {}&={}0,
    \end{split}
\end{equation}
where $\kappa = \dfrac{Nl^2}{2L^2}$. Denote the column vector $U=(\rho_1,\rho_2)^T$ and column of the  fluxes $F(U)=\frac{\sqrt{2}}{2}(\rho_1-\kappa\rho_1\rho_2,-\rho_2+\kappa\rho_1\rho_2)^T.$ Then the system of equations is expressed as
\[
\partial_t U + \grad_x F(U){}={}0.
\]
The system is hyperbolic if the eigenvalues of the gradient matrix 
\[
\grad_U F(U){}={}
\left[
\begin{array}{cc}
1-\kappa\rho_2 & -\kappa \rho_1\\
\kappa \rho_2 & -1+\kappa\rho_1
\end{array}\right]
\]
are real. The eigenvalues equal
\[
\lambda_\pm = \frac{1}{\sqrt{2}}\left(
\kappa(\rho_1-\rho_2) \pm \sqrt{\kappa^2(\rho_1-\rho_2)^2 + 4 - 4\kappa(\rho_1+\rho_2)}\right).
\]
Thus, the system is hyperbolic whenever
\[ 
\kappa^2(\rho_1-\rho_2)^2 + 4 - 4\kappa(\rho_1+\rho_2)\geq 0
\]
with a sufficient condition
\[
\kappa(\rho_1+\rho_2)\leq 1.
\]
Since \eqref{eq:kappa_bound} implies $\kappa < 0.5$, our kinetic model \eqref{Model4:2angles_sim} must be restricted to cases where the sum of the maximum density profiles for $\rho_1$ and $\rho_2$ total less than 2.

\section{Numerical Methods}\label{append:num}

 We express the system in \eqref{Model4:2angles_sim} in flux form
\begin{equation}
\partial_t U = - \partial_x F, 
\end{equation}
with $U = (\rho_1, \rho_2)^T$ and $F=A(\rho_1, \rho_2)^T$. In particular,
for two orientations $\pi/4$ and $3\pi/4$ the
fluxes become
$F_1 = \rho_1(1+\rho_2)$ and $F_2 = \rho_2(1+\rho_1)$.
Next, we discretize the equation in space and time using the Lax-Friedrichs method \cite{LeVeque2002}
\[
U_i^{n+1} = U_i^n - \frac{\Delta t}{\Delta x}
[\hat{F}_{i+1/2} - \hat{F}_{i-1/2}]
\]
and define 
\[
\hat{F}_{i+1/2} = \frac{F(U_{i+1}^n) + F(U_i^n)}{2} - \frac{K}{2}(U_{i+1}^n - U_i^n)
\]
where $K=\cos(\theta_1)(1+|\sin(\theta_2-\theta_1)|)$ is the upper bound on the speed of propagation given $\rho_1, \, \rho_2 <1$. Therefore, the CFL condition for the scheme above is $K\Delta t / \Delta x < 1$, and the additional diffusion introduced by the scheme is proportional to $D = K \Delta x / 2$. The scheme is first order in space and time, but we found that it was sufficient for our simulations.


\end{appendix}

\bibliographystyle{plain}
\bibliography{referencesMain}

\begin{thebibliography}{10}

\bibitem{balagam15b}
Rajesh Balagam and Oleg~A. Igoshin.
\newblock Mechanism for {{Collective Cell Alignment}} in {{Myxococcus}} xanthus
  {{Bacteria}}.
\newblock {\em PLOS Computational Biology}, 11(8):e1004474, August 2015.

\bibitem{balagam14b}
Rajesh Balagam, Douglas~B. Litwin, Fabian Czerwinski, Mingzhai Sun, Heidi~B.
  Kaplan, Joshua~W. Shaevitz, and Oleg~A. Igoshin.
\newblock Myxococcus xanthus {{Gliding Motors Are Elastically Coupled}} to the
  {{Substrate}} as {{Predicted}} by the {{Focal Adhesion Model}} of {{Gliding
  Motility}}.
\newblock {\em PLOS Computational Biology}, 10(5):e1003619, May 2014.

\bibitem{baskaran08a}
Aparna Baskaran and M.~Cristina Marchetti.
\newblock Enhanced {{Diffusion}} and {{Ordering}} of {{Self}}-{{Propelled
  Rods}}.
\newblock {\em Phys. Rev. Lett.}, 101(26):268101, December 2008.

\bibitem{baskaran08}
Aparna Baskaran and M.~Cristina Marchetti.
\newblock Hydrodynamics of self-propelled hard rods.
\newblock {\em Phys. Rev. E}, 77(1):011920, January 2008.

\bibitem{baskaran09}
Aparna Baskaran and M.~Cristina Marchetti.
\newblock Statistical mechanics and hydrodynamics of bacterial suspensions.
\newblock {\em PNAS}, 106(37):15567--15572, September 2009.

\bibitem{Bertin}
E.~Bertin, M.~Droz, and G.~Grégoire.
\newblock Boltzmann and hydrodynamic description for self-propelled particles.
\newblock {\em Phys Rev E Stat Nonlin Soft Matter Phys}, 74, 2006.

\bibitem{Bogoliubov1946}
N.~N. Bogoliubov.
\newblock Kinetic equations.
\newblock {\em Journal of Physics USSR}, 10:265–274, 1946.

\bibitem{bolley12}
Fran{\c c}ois Bolley, Jos{\'e}~A. Ca{\~n}izo, and Jos{\'e}~A. Carrillo.
\newblock Mean-field limit for the stochastic {{Vicsek}} model.
\newblock {\em Applied Mathematics Letters}, 25(3):339--343, March 2012.

\bibitem{CercignaniBook}
C.~Cercignani.
\newblock {\em The {Boltzmann} Equation and Its Applications}.
\newblock Springer-Verlag, New York, 1988.

\bibitem{chou_active_2015}
Yen-Liang Chou and Thomas Ihle.
\newblock Active matter beyond mean-field: {Ring}-kinetic theory for
  self-propelled particles.
\newblock {\em Physical Review E}, 91(2):022103, February 2015.
\newblock Publisher: American Physical Society.

\bibitem{degond17}
Pierre Degond, Angelika Manhart, and Hui Yu.
\newblock A continuum model for nematic alignment of self-propelled particles.
\newblock {\em Discrete \& Continuous Dynamical Systems - B}, 22(4):1295, 2017.

\bibitem{degond18}
Pierre Degond, Angelika Manhart, and Hui Yu.
\newblock An age-structured continuum model for myxobacteria.
\newblock {\em Math. Models Methods Appl. Sci.}, 28(09):1737--1770, August
  2018.

\bibitem{degond08}
Pierre Degond and S{\'e}bastien Motsch.
\newblock Continuum limit of self-driven particles with orientation
  interaction.
\newblock {\em Math. Models Methods Appl. Sci.}, 18(supp01):1193--1215, August
  2008.

\bibitem{degond10}
Pierre Degond and Tong Yang.
\newblock Diffusion in a continuum model of self-propelled particles with
  alignment interaction.
\newblock {\em Math. Models Methods Appl. Sci.}, 20(supp01):1459--1490,
  September 2010.

\bibitem{Ginelli2010}
F.~Ginelli, F.~Peruani, M.~B\"{a}r, and H.~Chat\'{e}.
\newblock Large-scale collective properties of self-propelled rods.
\newblock {\em Phys Rev Lett}, 104, 2010.

\bibitem{Grad1949}
Harold Grad.
\newblock On the kinetic theory of rarefied gases.
\newblock {\em Communications on pure and applied mathematics}, 2:331–407,
  1949.

\bibitem{hanke_understanding_2013}
Timo Hanke, Christoph~A. Weber, and Erwin Frey.
\newblock Understanding collective dynamics of soft active colloids by binary
  scattering.
\newblock {\em Physical Review E}, 88(5):052309, November 2013.
\newblock Publisher: American Physical Society.

\bibitem{Manhart1}
Sabine Hittmeir, Laura Kanzler, Angelika Manhart, and Christian Schmeiser.
\newblock Kinetic modelling of colonies of myxobacteria.
\newblock {\em Kinetic and Related Models}, 14(1):1--24, 2021.

\bibitem{ihle_towards_2014}
T.~Ihle.
\newblock Towards a quantitative kinetic theory of polar active matter.
\newblock {\em The European Physical Journal Special Topics},
  223(7):1293--1314, June 2014.

\bibitem{ihle_kinetic_2011}
Thomas Ihle.
\newblock Kinetic theory of flocking: {Derivation} of hydrodynamic equations.
\newblock {\em Physical Review E}, 83(3):030901, March 2011.
\newblock Publisher: American Physical Society.

\bibitem{jeggle20}
Julian Jeggle, Joakim Stenhammar, and Raphael Wittkowski.
\newblock Pair-distribution function of active brownian spheres in two spatial
  dimensions: Simulation results and analytic representation.
\newblock {\em The Journal of Chemical Physics}, 152(19):194903, 2020.

\bibitem{kaiser03}
Dale Kaiser.
\newblock Coupling cell movement to multicellular development in myxobacteria.
\newblock {\em Nat Rev Microbiol}, 1(1):45--54, October 2003.

\bibitem{kursten21}
Rüdiger Kürsten and Thomas Ihle.
\newblock Quantitative kinetic theory of flocking with three-particle closure.
\newblock {\em Physical Review E}, 104(3):034604, 2020.

\bibitem{kursten20}
Rüdiger Kürsten, Sven Stroteich, Martín~Zumaya Hernández, and Thomas Ihle.
\newblock Multiple {{Particle Correlation Analysis}} of {{Many-Particle
  Systems}}: {{Formalism}} and {{Application}} to {{Active Matter}}.
\newblock {\em Physical Review Letters}, 124(8):088002, 2020.

\bibitem{LeVeque2002}
R.~LeVeque.
\newblock {\em Finite volume methods for hyperbolic problems}.
\newblock Cambridge University Press, 2002.

\bibitem{narayan07}
Vijay Narayan, Sriram Ramaswamy, and Narayanan Menon.
\newblock Long-{{Lived Giant Number Fluctuations}} in a {{Swarming Granular
  Nematic}}.
\newblock {\em Science}, 317(5834):105--108, July 2007.

\bibitem{ndlec97}
F.~J. Ndlec, T.~Surrey, A.~C. Maggs, and S.~Leibler.
\newblock Self-organization of microtubules and motors.
\newblock {\em Nature}, 389(6648):305--308, September 1997.

\bibitem{perepelitsa_mean-field_2022}
Misha Perepelitsa, Ilya Timofeyev, Patrick Murphy, and Oleg~A. Igoshin.
\newblock Mean-field model for nematic alignment of self-propelled rods.
\newblock {\em Physical Review E}, 106(3):034613, September 2022.
\newblock Publisher: American Physical Society.

\bibitem{peruani10}
F.~Peruani, L.~{Schimansky-Geier}, and M.~B{\"a}r.
\newblock Cluster dynamics and cluster size distributions in systems of
  self-propelled particles.
\newblock {\em Eur. Phys. J. Spec. Top.}, 191(1):173--185, December 2010.

\bibitem{peruani06a}
Fernando Peruani, Andreas Deutsch, and Markus B{\"a}r.
\newblock Nonequilibrium clustering of self-propelled rods.
\newblock {\em Phys. Rev. E}, 74(3):030904, September 2006.

\bibitem{Peruani2008}
Fernando Peruani, Andreas Deutsch, and Markus B{\"a}r.
\newblock A mean-field theory for self-propelled particles interacting by
  velocity alignment mechanisms.
\newblock {\em Eur. Phys. J. Spec. Top.}, 157:111--122, 2008.

\bibitem{peruani_collective_2012}
Fernando Peruani, Jörn Starruß, Vladimir Jakovljevic, Lotte
  Søgaard-Andersen, Andreas Deutsch, and Markus Bär.
\newblock Collective {Motion} and {Nonequilibrium} {Cluster} {Formation} in
  {Colonies} of {Gliding} {Bacteria}.
\newblock {\em Physical Review Letters}, 108(9):098102, February 2012.
\newblock Publisher: American Physical Society.

\bibitem{peshkov_boltzmann-ginzburg-landau_2014}
A.~Peshkov, E.~Bertin, F.~Ginelli, and H.~Chaté.
\newblock Boltzmann-{Ginzburg}-{Landau} approach for continuous descriptions of
  generic {Vicsek}-like models.
\newblock {\em The European Physical Journal Special Topics},
  223(7):1315--1344, June 2014.

\bibitem{pismen_active_2021}
Len Pismen.
\newblock {\em Active {Matter} {Within} and {Around} {Us}: {From}
  {Self}-{Propelled} {Particles} to {Flocks} and {Living} {Forms}}.
\newblock The {Frontiers} {Collection}. Springer International Publishing,
  Cham, 2021.

\bibitem{riedel05}
Ingmar~H. Riedel, Karsten Kruse, and Jonathon Howard.
\newblock A {{Self}}-{{Organized Vortex Array}} of {{Hydrodynamically Entrained
  Sperm Cells}}.
\newblock {\em Science}, 309(5732):300--303, July 2005.

\bibitem{schaller10}
Volker Schaller, Christoph Weber, Christine Semmrich, Erwin Frey, and
  Andreas~R. Bausch.
\newblock Polar patterns of driven filaments.
\newblock {\em Nature}, 467(7311):73--77, September 2010.

\bibitem{sumino12}
Yutaka Sumino, Ken~H. Nagai, Yuji Shitaka, Dan Tanaka, Kenichi Yoshikawa,
  Hugues Chat{\'e}, and Kazuhiro Oiwa.
\newblock Large-scale vortex lattice emerging from collectively moving
  microtubules.
\newblock {\em Nature}, 483(7390):448--452, March 2012.

\bibitem{suzuki_polar_2015}
Ryo Suzuki, Christoph~A. Weber, Erwin Frey, and Andreas~R. Bausch.
\newblock Polar pattern formation in driven filament systems requires
  non-binary particle collisions.
\newblock {\em Nature Physics}, 11(10):839--843, October 2015.
\newblock Number: 10 Publisher: Nature Publishing Group.

\bibitem{thutupalli15c}
Shashi Thutupalli, Mingzhai Sun, Filiz Bunyak, Kannappan Palaniappan, and
  Joshua~W. Shaevitz.
\newblock Directional reversals enable {{Myxococcus}} xanthus cells to produce
  collective one-dimensional streams during fruiting-body formation.
\newblock {\em J R Soc Interface}, 12(109):20150049, August 2015.

\bibitem{thuroff_critical_2013}
Florian Thüroff, Christoph~A. Weber, and Erwin Frey.
\newblock Critical {Assessment} of the {Boltzmann} {Approach} to {Active}
  {Systems}.
\newblock {\em Physical Review Letters}, 111(19):190601, November 2013.
\newblock Publisher: American Physical Society.

\bibitem{weber_nucleation-induced_2012}
Christoph~A. Weber, Volker Schaller, Andreas~R. Bausch, and Erwin Frey.
\newblock Nucleation-induced transition to collective motion in active systems.
\newblock {\em Physical Review E}, 86(3):030901, September 2012.
\newblock Publisher: American Physical Society.

\bibitem{weber_role_2013}
Christoph~A. Weber, Florian Thüroff, and Erwin Frey.
\newblock Role of particle conservation in self-propelled particle systems.
\newblock {\em New Journal of Physics}, 15(4):045014, April 2013.
\newblock Publisher: IOP Publishing.

\bibitem{wu07}
Yilin Wu, Yi~Jiang, Dale Kaiser, and Mark Alber.
\newblock Social {{Interactions}} in {{Myxobacterial Swarming}}.
\newblock {\em PLoS Comput Biol}, 3(12):e253, December 2007.

\end{thebibliography}

\end{document}